\documentclass[journal]{IEEEtran}
\usepackage{cite}

\ifCLASSINFOpdf
   \usepackage[pdftex]{graphicx}
	\usepackage{epstopdf}
\else
   \usepackage[dvips]{graphicx}
\fi
\usepackage[cmex10]{amsmath}
\ifCLASSOPTIONcompsoc
  \usepackage[caption=false,font=normalsize,labelfont=sf,textfont=sf]{subfig}
\else
  \usepackage[caption=false,font=footnotesize]{subfig}
\fi

\usepackage{siunitx}
\begin{document}
%
\title{Electron Back Scattering in CNTFETs}
%
%
%

\author{Igor~Bejenari~
        and~Martin~Claus 
\thanks{Manuscript received October xx, 2015; revised December xx, xx. This work was supported in part by a grant from the Cfaed and  DFG project CL384/2.}%
\thanks{I. Bejenari and M. Claus are with the Chair for Electron Devices and Integrated Circuits, Department of Electrical and Computer Engineering, 
 Technische Universit\"at Dresden, 01062, Germany.}%
\thanks{I. Bejenari is  also with Institute of Electronic Engineering and Nanotechnologies,  Academy of Sciences of Moldova, MD 2028 Chisinau, Moldova (e-mail:igor.bejenari@fulbrightmail.org).}
\thanks{M. Claus is also  with the Center for Advancing Electronics Dresden (Cfaed),
 Technische Universit\"at Dresden, 01062, Germany (e-mail:Martin.Claus@tu-dresden.de).}
}%
%
%

\markboth{IEEE TRANSACTIONS ON ELECTRON DEVICES,~Vol.~x, No.~xx, December~xx}%
{Bejenari \MakeLowercase{\textit{et al.}}: Electron Back Scattering in CNTFETs}
%



\IEEEoverridecommandlockouts
 \IEEEpubid{\makebox[\columnwidth]{978-1-4799-7492-4/15/\$31.00~
 \copyright2015
 IEEE \hfill} \hspace{\columnsep}\makebox[\columnwidth]{ }}

\maketitle

\begin{abstract}
A new  non-ballistic analytical model for the intrinsic channel region of MOSFET-like single-walled carbon-nanotube field-effect transistors with ohmic contacts has been developed which overcomes the limitations of existing models and extends their applicability toward high bias voltages needed for analog applications. The new model comprises an improved description of electron-phonon scattering mechanism taking into account the accumulation of electrons at the bottom of conduction subband due to back scattering by optical phonons. The model has been justified by a Boltzmann transport equation solver. The simulation results are found to be in agreement with experimental data for highly doped CNTFETs.
\end{abstract}


%
\IEEEpeerreviewmaketitle

\section{Introduction}
%
%
%
%

\IEEEPARstart{T}{he} carbon-nanotube field-effect transistor (CNTFET) represents a potential candidate to compete traditional silicon MOSFETs especially for analog high-frequency applications~\cite{Schroter_2013}.
In a high-quality intrinsic carbon nanotube (CNT), the electron-electron scattering is negligible and the dominant scattering mechanism  is electron-phonon scattering~\cite{Yao2000, Mann2003,Javey2004, Park2004}. In CNTs under low voltage bias, the electron mean free path (mfp) is observed to be very long (\SI{1}{\micro\meter}), and is supposed to be nearly elastic and limited by acoustic phonon scattering \cite{Mann2003}. Under high bias, optical phonon emission dominates and this results in a rather short  electron mfp of about \SI{10}{\nm}. Due to the large optical phonon energy (${\hbar \omega_{op} \approx \SI{0.16}{\eV}}$) in CNTs, the DC current is near-ballistic at low gate biases even though a significant amount of scattering exists near the drain end of the CNT channel~\cite{Guo2005}. However, under high biases, optical phonon scattering dominates the transport along the whole channel.

Early p-type CNTFETs were fabricated by using Pt and Au contacts~\cite{Tans1998,Martel1998}. Transport in these devices is dominated by the Schottky barriers (SB) existing at the metal source/drain contacts~\cite{Heinze2002}. As a result, the ON-state current is reduced~\cite{Chen2005}. Since the RC time constant associated with the contact resistance can limit the frequency response of the devices, ohmic contacts with a low resistivity  are preferable for high-frequency operation. Different technologies have been developed to form n- and p-type ohmic contacts in CNTFETs. While the palladium (Pd) contacts  have been demonstrated to be  barrier-free for hole carrier transport, gadolinium (Gd), yttrium  (Y) and scandium (Sc) form n-type ohmic contacts with CNTs~\cite{Chen2005,Javey2005,Wang2011,Ding2009,Zhang2007,Zhang2008}. To further improve the contact transparency, different doping techniques have been developed~\cite{Appenzeller2004}.
 
The model execution time is a main stumbling block in the development of a circuit-level model. Therefore, compact models are the preferred simulation tool for circuit designers to assess the actual performance potential of a technology. As it will be discussed below, the physics-based compact models reported in literature~\cite{Wong_IEEE2007,Wong_IEEE2007PII,Fregonese_IEEE2009,Kazmierski_IEEE2010} predict an unphysical CNTFET behavior at high biases. These models are, thus, not suitable for the design of analog high-frequency applications.

In this paper, we propose a physics-based analytical model for the electron transport through the intrinsic channel of a CNTFET with ohmic contacts. For the sake of simplicity, the transport is evaluated in the quantum capacitance limit (QCL). Although a complete compact model suitable for circuit design should take into account the effect of the non-equilibrium mobile charge and terminal charges on the electrostatics, they are not considered here~\cite{WCM_2012}. However, the proposed analytical models can be easily included in the compact models, which consider both the electron transmission and charge distribution in the channel.
Without loss of generality, we study CNTFETs with n-type ohmic contacts and consider  electron scattering by both acoustic and optical phonons.  The channel length is set to \SI{100}{\nano\meter} which is close to the sweet spot in silicon CMOS for high-frequency analog applications. 
 \IEEEpubidadjcol 

\section{CNTFET transport model}
\label{sec:level2} 

We consider two physics-based approaches for current calculations. In the first approach, the transmission probability is a step-like function with respect to energy. In the other approach, the transmission probability is a continuous energy-dependent function.

The net electric current $I$ equals to the current $I_S$ flowing from the source to the drain 
(${+k}$ branch) minus the current $I_D$ flowing from the drain to the source (${-−k}$ branch)~\cite{Datta_1995}
\begin{equation}
I = \frac{4q}{h} \sum_{m} {\int\limits_{E^c_{m}(x_{cc})}^\infty \left[ f^{+}_{m}(E) - f^{-}_{m}(E) \right]dE}=I_{S}-I_{D},
\label{eq:one1}
\end{equation}
where ${ f^{\pm}_{m}(E)}$  denotes a non-equilibrium distribution function of electrons with positive and negative momenta ${\pm k}$, respectively, and corresponding to the $m$th  electron subband $E_m(k)$ at the current control point  ${x_{cc}}$ in the CNT channel.
 The current control point  is estimated at the bottom of the space-dependent conduction subband ${E^c_{m}(x)}$ using the condition that the transmission  through that subband becomes negligible for electrons with energy less than the current control energy ${E^c_{m}(x_{cc})}$~\cite{Mothes2015}.
The last represents the lower limit of integration. 
The similar concept of the reduction of conduction band profile to a single current control point was used for Si MOSFETs~\cite{Natori94,Khakifirooz_IEEE2009}. 
Here, we suppose the 1st electron subband edge $E_1(0)$ to be a half of the CNT band gap $E_g$.
For CNTs, the product of  the spin and electron subband degeneracies gives a factor of 4 in front of the integral in~(\ref{eq:one1}).

The non-equilibrium distribution functions $f^{+}$ and $f^{-}$ in~(\ref{eq:one1}) can be obtained as a solution of the Boltzmann transport equation (BTE) by using the Monte Carlo method~\cite{Lundstrom_2000,Mothes2015}.
In compact models, ${ f^{\pm}(E)}$ is usually calculated  in the framework of the Landauer formalism~\cite{Wong_IEEE2007}
\begin{eqnarray}
f^{+}(E) = T_{LR}(E,V_{DS}) f_{0} \left(E-q\psi_{cc} \right),
\\
f^{-}(E) = T_{RL}(E,0) f_{0} \left(E-q\psi_{cc}+qV_{DS} \right)
\label{eq:two2},
\end{eqnarray}
where ${V_{DS}}$ is a drain-source voltage, ${T_{LR}(E,V_{DS})}$  is a transmission probability of electrons to propagate from the source (left contact)  to the drain (right contact), ${\psi_{cc}}$ is electrostatic (tube) potential defined at the current control point ${x_{cc}}$.
The source and drain regions are assumed to be in thermodynamic equilibrium with the metal contacts in the CNTFETs. Therefore, the contact carrier statistics is calculated by means of  the Fermi-Dirac distribution function $f_{0}$.

In the general case, the transmission probability of electrons continuously depends on energy as~\cite{Wong_IEEE2007} 
\begin{equation}
{T(E,V_{xs})}=\frac{l_{eff}(E,V_{xs})}{l_{eff}(E,V_{xs})+L_{g}},
\label{eq:Teff}
\end{equation}
where ${L_{g}}$ is the CNT channel length,  ${V_{xs}}$ is the potential drop between the node $x$ along the channel and the source, and ${l_{eff}}$ is the effective mfp of electrons.
For electrons scattered by both acoustic and optical phonons, ${l_{eff}}$ is defined by means of the Mathiessen law as ${l^{-1}_{eff}(E,V_{xc})=l^{-1}_{ac}(E,V_{xc})+l^{-1}_{op}(E,V_{xc})}$. 
In this expression, the effective mfp ${l_{ac}}$ of electrons scattered by acoustic phonons is defined as
\begin{equation}
l_{ac}(E,V_{xs})=\frac{D_{0}}{D(E)}\frac{\lambda_{ac}}{\left[1-f_{0} \left(E-q\psi_{cc}+qV_{xs} \right)\right]},
\label{eq:lac}
\end{equation}
where ${D(E)}$ is the 1D electron density of states. Factor ${\left[1-f_{0} \left(E-q\psi_{cc}+qV_{xs} \right)\right]}$ was introduced to estimate a probability of electron backscattering taking into consideration Pauli's exclusion principle.
The high energy electrons can be back scattered by emission of optical phonons with an energy of ${\hbar\omega_{op}\approx\SI{0.16}{\eV}}$~\cite{Javey2004}.
In this case, the electron effective mfp ${l_{op}}$ is given by
\setlength{\arraycolsep}{0.0em}
\begin{eqnarray}
l_{op}(E,V_{xs})&{}={}&{D_{0}}/{D(E-\hbar \omega_{op})}
\nonumber\\
&&\times
 \frac{\lambda_{op}}{\left[1-f_{0} \left(E-\hbar \omega_{op}-q\psi_{cc}+qV_{xs} \right)\right]}.
\label{eq:lop}
\end{eqnarray}
\setlength{\arraycolsep}{5pt}

In a simplified model, the probability for electrons depending on energy has only two different constant values ${T_{ac}}$ and ${T_{high}}$, which are defined by~(\ref{eq:Teff}) with modified $l_{eff}$~\cite{Fregonese_IEEE2009}.
The transmission probability ${T_{ac}}$ of low energy electrons scattered only by acoustic phonons  is defined in terms of  ${l_{eff}=l_{ac}=\lambda_{ac}/\sqrt{1+qE_{g}/(\alpha k_{B}T)}}$.
Parameter $\alpha$ is equal to 1.83 for a range of temperature from 200 K to 400 K and for a range of chirality from (13,0) to (25,0)~\cite{Fregonese_IEEE2009}.
In this approach, the electron density of states is replaced by its average value and  Pauli's exclusion principle is neglected.
For high energy electrons scattered by both acoustic and optical phonons, the transmission probability, ${T_{high}}$, is defined in terms of ${l_{eff}=(l^{-1}_{ac}+\lambda^{-1}_{op}})^{-1}$, where ${\lambda_{op}}$ is mfp of electrons scattered by optical phonons. 

Below, we restrict our study to QCL, when the tube potential $\psi_{cc}$ is proportional to the applied gate voltage ${V_{g}}$. Also, the conduction subband edges are assumed to be constant along the CNT channel. The electron scattering by acoustic and optical phonons is included both within the BTE solver and with all considered approximations. Because of the limited  number of optical phonons with an energy of \SI{0.16}{\eV} at room temperature, we do not consider the absorption of optical phonons by electrons. 

\section{Limitation of existing Compact Models}
\label{sec:level3} 

\subsection{Model A}
\label{sec:level32} 

Using the step-like approximation for the transmission probability as discussed in the above mentioned simplified model, the integral in~(\ref{eq:one1}) can be solved analytically. 
In this model, the source and drain components of electric current read~\cite{Fregonese_IEEE2009}
\begin{eqnarray}
&&{I_{S(D)}}=\frac{4qk_B T}{h}
\nonumber\\
&& \times\sum_{m}\left\{ T_{ac} \ln\left[1+ \exp\left(\frac{q(\psi_{cc}-V_{S(D)})-E_m(0)}{k_B T}\right)\right]   \right.
\nonumber\\
&& +\left.  \ln\left[1+ \exp\left(\frac{q(\psi_{cc}-V_{S(D)})-E_m(0)-\hbar\omega_{op}}{k_B T}\right)\right]  \right.  \nonumber\\
&& \times \left.    (T_{high}-T_{ac})  \right\},
\label{eq:I_S_Freg}
\end{eqnarray} 
where $m$ is an index of electron subband ${E_{m}(k)}$.

This approach has successfully been applied to obtain the current-voltage (I-V) characteristics at a low gate voltage, when the electrons injected only from source contact contribute to the net current.  
However, at a high gate bias, this model predicts an unphysical behavior of the saturation current.

\begin{figure}[!t]
\centering
\includegraphics[width=8cm]{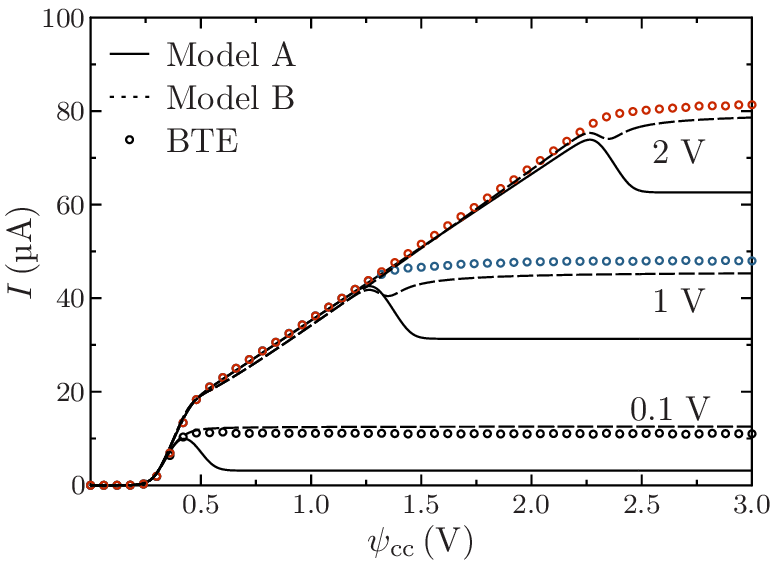}
\caption{(Color online) The net current ${I_{\text{DS}}}$ calculated in the models A, B and BTE approximation as a function of tube potential $\psi_{cc}$ at drain-source voltage ${V_{DS}}$ equal to  0.1, 1.0, and 2.0 V. CNT chirality (19,0), band gap ${E_g=0.579}$ eV, gate length ${L_{g}=100}$ nm, temperature ${T=300}$ K.}
\label{fig:1}
\end{figure}

Figure~\ref{fig:1} depicts the net current $I$ as a function of tube potential $\psi_{cc}$ for different drain-source voltages taking into account the contribution of the first electron subband.
The current obtained in the framework of the BTE theory is also presented for comparison.                        
In contrast to the BTE results, the current calculated by means of the analytical approach suddenly drops at the onset of saturation leading to an unphysical negative transconductance.

The drop in the current can be explained by evaluating~(\ref{eq:I_S_Freg}) at two bias points: (\textit{i}) at the onset of saturation, where the tube potential, $q\psi_{cc}$, equals $q\psi_{cc,1}=qV_{DS}+E_{1}(0)$, and (\textit{ii}) far beyond the onset of saturation, where the tube potential is equal to or greater than $q\psi_{cc,2}=qV_{DS}+E_{1}(0) + \hbar\omega_{op} +3 k_{B}T$.

\begin{figure}[!t]
\centering
\includegraphics[width=7cm]{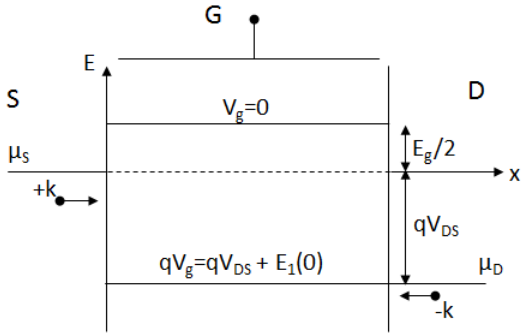}
\caption{\label{fig:2} The conventional schematic band diagram of CNTFET in the quantum capacitance limit (${\psi_{cc} \approx V_{g}}$). The chemical potential ${\mu_{S}}$ of the source contact is set in the middle of the CNT band gap ${E_g}$.}
\end{figure}
At $\psi_{cc}=\psi_{cc,1}$, the bottom of the conduction subband approaches the chemical potential (Fermi level) $\mu_{D}$  (see Fig.~\ref{fig:2}) which allows electron injection from the drain and, thus, a saturation of the net current by compensating source injected electrons.    
For ${qV_{DS} \geq \hbar\omega_{op} + 3 k_{B}T}$,  the saturation current given by~(\ref{eq:I_S_Freg})  simplifies to
\begin{equation}
  \label{eq:I_sat1}  
  I_{sat,1}\approx(4q^2/h)T_{high}V_{DS}+(4q/h)[T_{ac}-T_{high}]\hbar\omega_{op}
\end{equation}
where the last term takes into consideration the electron back scattering by optical phonons. 
However, at $\psi_{cc,2}$, i.e., far beyond the onset of saturation, the saturation current based on~(\ref{eq:I_S_Freg}) reduces to 
\begin{equation} 
  I_{sat,2}=I_{S}-I_{D}\approx (4q^2/h)T_{high}V_{DS}.
	  \label{I_sat2}
\end{equation}
The missing electron back scattering in the last expression leads to the unphysical current drop in the saturation region of the current. 

\subsection{Model B}
\label{sec:level33}  

In this approach, the transmission probability is the continuous energy-dependent function given by~(\ref{eq:Teff}) along with~(\ref{eq:lac}) and~(\ref{eq:lop}). The integration in~(\ref{eq:one1}) is replaced by a summation over discrete longitudinal wave vector of electrons. This significantly simplifies the calculation of the net current~\cite{Wong_IEEE2007}.
The method is very useful for short gate length devices, if the wave vector discretization is large.  

Along with the model A, this approach successfully describes the I-V characteristics at a low gate bias, but it fails at a high gate bias as shown in Fig.~\ref{fig:1}.  
The net current shows unphysical dips in the current at the onset of saturation, because the source component of the current is underestimated, whereas the drain component of the current is overestimated in this region. In contrast to the model A, the transmission probability depends on both the gate and source-drain bias. Therefore, with further increase of the tube potential, the current approaches the value calculated by using the BTE method.

\section{Optical-phonon-induced back scattering}
\label{sec:level4} 

\begin{figure}[!t]
\centering
\includegraphics[width=8cm]{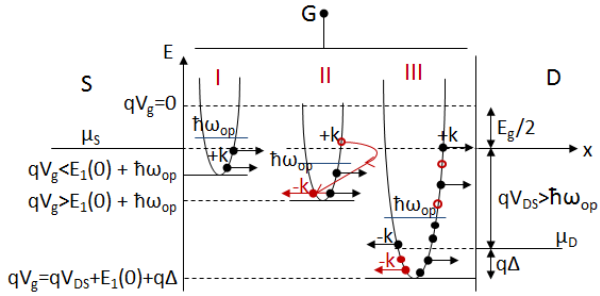}
\caption{(Color online) The schematic band diagram of CNTFET in the quantum capacitance limit taking into account the  electron back scatting by emitting optical phonons.}
\label{Fig:BandDiagram2}
\end{figure}

Fig.~\ref{Fig:BandDiagram2} depicts the schematic band diagram of the CNTFET taking into account the electron back scattering by optical phonons. Three different gate bias conditions are shown.  The electron injection from the drain into the channel is allowed only in case of the third (highest) gate bias.  At ${qV_{g}<E_{1}(0)+\hbar\omega_{op}}$  (bias condition I) or ${qV_{DS}<\hbar\omega_{op}}$, the electrons injected into the channel from the source contact can be scattered only by acoustic phonons, if any. In this case, the electron energy is conserved. At ${qV_{g}>E_{1}(0)+\hbar\omega_{op}}$ and ${qV_{DS}>\hbar\omega_{op}}$ (bias condition II), high energy electrons can be scattered backward by emitting an optical phonon, i.e, changing their momentum ${+k}$ to opposite one ${-k}$ and losing their energy. After scattering, these electrons mostly occupy states at the bottom of the conduction subband.
Hence, when the conduction subband edge approaches the chemical potential, ${\mu_{D}}$, of the drain contact by increasing the gate voltage to ${qV_{g}=qV_{DS}+E_{1}(0)}$, the electrons can not be injected into the channel from the drain in accordance with Pauli\text{'}s exclusion principle, because the low energy electron states with the negative momentum ${-k}$ have been already occupied by backscattered source injected electrons.
To open the channel for the drain electrons, one has to increase the gate voltage by a small amount of ${\Delta}$ (bias condition III), corresponding to the upper energy of the back scattered electrons accumulated at the bottom of the conduction subband (see Fig.~\ref{Fig:BandDiagram2}).
Thus, the transmission probability of drain electrons is reduced once backscattered source injected electrons block the injection of electrons from the drain contact. 

Therefore, a detailed study of scattering in CNTs revealed that the injection of drain electrons into the channel is affected by back scattered source electrons accumulated at the bottom of the conduction subband, a phenomenon which has not been considered so far in the approaches discussed in the literature.   

\section{Modified model A}%

In this section, we introduce a new physics-based analytical model for the net current.
If electrons are scattered only by acoustic phonons, then the source and drain components of the current are given by
\setlength{\arraycolsep}{0.0em}
\begin{eqnarray}
&&{I_{S(D)}}=\frac{4qk_B T}{h}
\nonumber\\
&& \times\sum_{m}\left\{T_{ac} \ln\left[1+\exp \left(\frac{q(\psi_{cc}-V_{S(D)})-E_m(0)}{k_B T}\right)\right]\right\}.
\nonumber\\
\label{eq:I_SD}
\end{eqnarray}
\setlength{\arraycolsep}{5pt} 
In case electrons are scattered by both acoustic and optical phonons, the source and drain components of current are defined as
\setlength{\arraycolsep}{0.0em}
\begin{eqnarray}
&&{I_{S}}=\frac{4qk_B T}{h}
\nonumber\\
&&\sum_{m}\left\{(T_{high}-T_{ac}) \ln\left[1+ \exp\left(\frac{q\psi_{cc}-E_m(0)-\hbar\omega_{op}}{k_B T}\right)\right]\right.
\nonumber\\
&& +\left.T_{ac} \ln\left[1+ \exp\left(\frac{q\psi_{cc}-E_m(0)}{k_B T}\right)\right]\right\},
\label{eq:I_S_op}
\end{eqnarray} 
\begin{eqnarray}
&&{I_{D}}=\frac{4qk_B T}{h}
\nonumber\\
&& \sum_{m}\left\{T_{high} \ln\left[1+ \exp\left(\frac{q(\psi_{cc}-V_{DS}-\Delta)-E_m(0)}{k_B T}\right)\right]\right\}.
\nonumber\\
\label{eq:I_D_op}
\end{eqnarray} 
\setlength{\arraycolsep}{5pt}
Using~(\ref{eq:I_S_op}) and (\ref{eq:I_D_op}), one can deduce that the net saturation current ${I_{sat,2}=I_{S}(\psi_{cc,2})-I_{D}(\psi_{cc,2})}$ is in agreement with~(\ref{eq:I_sat1}) after replacing ${V_{DS}}$ by ${V_{DS}+\Delta}$. In this model, we have obtained by a fitting procedure a linear dependence of the parameter ${\Delta}$ on the applied source-drain voltage,i.e., ${\Delta=0.1V_{DS}}$. 

\section{Modified model B}%

In this section, we improve the model B taking into consideration the accumulation of backscattered source electrons at the bottom of the conduction subband.
Using the approach described in Section~\ref{sec:level33},  we can express the net current based on~(\ref{eq:one1}) as
\begin{eqnarray}
I&=&2\sum_{m,l}\left[T_{LR}(E_m(k_l),V_{DS}-\Delta) J_{m,l}(0,\psi_{cc}) \right.
\nonumber\\
&&  \left.- T_{RL}(E_m(k_l),0) J_{m,l}(V_{DS}+\Delta,\psi_{cc}) \right],
\label{eq:I}
\end{eqnarray}
Here, the parameter ${\Delta}$ is of order of the electron thermal energy ${k_{B}T}$ in case  electrons are scattered by both acoustic and optical phonons.
If electrons are scattered only by acoustic phonons, then the effective mfp ${l_{eff}\rightarrow l_{ac}}$  and parameter ${\Delta}$ tends to zero.
Introducing the parameter ${\Delta}$, we diminish the underestimation (overestimation) of the source (drain) component of current and exclude the unphysical dip in function ${I(\psi_{cc})}$.
The accumulation of back scattered electrons at the bottom of the conduction subband leads to a decrease of the scattering of electrons injected from the source contact at ${\psi_{cc}=E_{1}(0)+V_{DS}-q\Delta}$ before the electrons start to inject into the channel from the drain contact at ${q\psi_{cc}=E_{1}(0)+qV_{DS}+q\Delta}$.

\section{Results}
\label{sec:level7}  

Here, we compare transport characteristics of CNTFETs  obtained in the modified models A and B with those calculated by using the BTE solver. Also, we compare our results with available experimental data. The electron scattering parameters used for the calculations are listed in Table~\ref{tab:table1}~\cite{Fregonese_IEEE2009}. In contrast to the model A, mfp ${\lambda_{ac}=450}$ nm is about twice less than ${\lambda_{ac}=963}$ nm used in the BTE calculations. In both models A and B, mfp ${\lambda_{op}}$ is about twice greater than ${\lambda_{op}=15}$ nm used in the BTE model. 
\begin{table}[!t]
\renewcommand{\arraystretch}{1.3}
\caption{Electron scattering parameters used for the calculations}
\label{tab:table1}
\centering
\begin{tabular}{|c||c|c |c|}
\hline
 & ${\lambda_{ac}(\SI{}{\nano\meter})}$ & ${\lambda_{op}(\SI{}{\nano\meter})}$ & ${\Delta}(\SI{}{\V})$\\
\hline
Model A  & 963 & 28 &  ${0.1 V_{DS}}$\\
\hline
Model B  & 450 & 30 & ${0.75k_{B}T/q}$\\
\hline
BTE solver & 963  & 15 & --\\
\hline
\end{tabular}
\end{table}

\begin{figure}[!t]
\centering
\includegraphics[width=8cm]{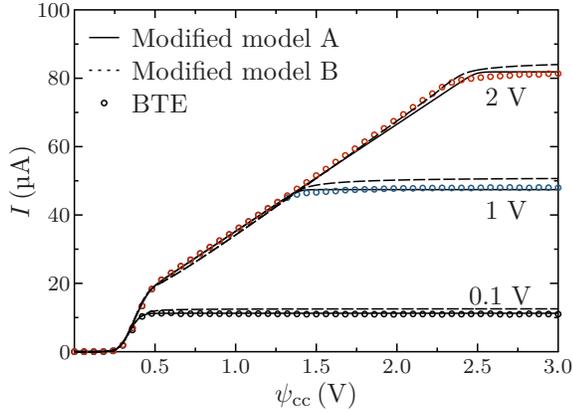}
\caption{(Color online) The net current ${I}$ calculated in the modified models A, B and BTE approximation as a function of tube potential $\psi_{cc}$ at drain-source voltage ${V_{DS}}$ equal to  0.1, 1.0, and 2.0 V. CNT chirality (19,0), band gap ${E_g=0.579}$ eV, gate length ${L_{g}=100}$ nm, temperature ${T=300}$ K.}
\label{fig:5}
\end{figure}

Fig.~\ref{fig:5} shows the net current ${I}$ as a function of tube potential $\psi_{cc}$ at different values of drain-source voltage ${V_{DS}}$ in the modified models A and B for CNTFET with a gate length of \SI{100}{\nano\meter}.
The results obtained in these models agree with the BTE calculations.
In this case, the drop at high  $\psi_{cc}$ in the dependence of ${I(\psi_{cc})}$ is eliminated.
At ${q\psi_{cc}\approx E_{1}(0)}$, the bottom of the conduction subband approaches the chemical potential ${\mu_{S}}$ of the source contact and electrons start to flow from the source to drain.
The slope of ${I(\psi_{cc})}$ in the interval ${E_{1}(0)<q\psi_{cc}<E_{1}(0)+\hbar \omega_{op}}$ is mainly defined by the mfp ${\lambda_{ac}}$ of electrons scattered by only acoustic phonons.
Due to scattering of electrons by both acoustic and optical phonons, the slope of ${I(\psi_{cc})}$ decreases in the interval ${E_{1}(0)+\hbar \omega_{op}<q\psi_{cc}<E_{1}(0)+qV_{DS}+q\Delta}$.
At ${q\psi_{cc} \approx E_{1}(0)+qV_{DS}+q\Delta}$, electrons are injected into the channel from the drain contact. These partially compensate the source electrons and the net current achieves constant value ${I_{sat}}$, which is not changed with a further increase of the tube potential ${\psi_{cc}}$. 
At ${q\psi_{cc} \approx E_{1}(0)+qV_{DS}}$,  the dip in the dependence of ${I(\psi_{cc})}$ is eliminated.

\begin{figure}[!t]
\centering
\includegraphics[width=8cm]{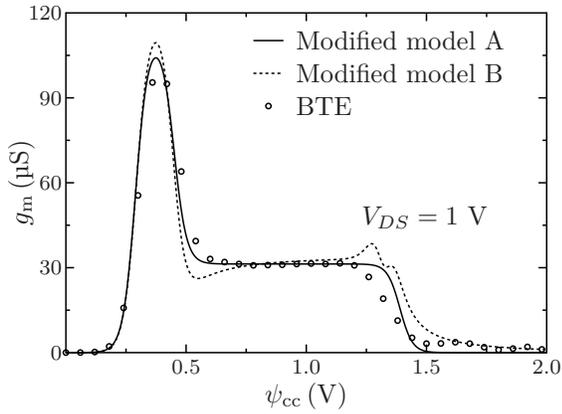}
\caption{The transconductance ${g_{m}}$ calculated in the BTE approach, modified  models A and B as a function of tube potential $\psi_{cc}$ at drain-source voltage ${V_{DS}=1}$ V. CNT chirality (19,0),  band gap ${E_g=0.579}$ eV, gate length ${L_{g}=100}$ nm, and temperature ${T=300}$ K.}
\label{fig:6} 
\end{figure}

Fig.~\ref{fig:6} compares the transconductances, ${g_{m}=\partial I/\partial \psi_{cc}}$, calculated in the  modified models A and B, as well as with the BTE approach.
The maximum of ${g_{m}}$ is located at ${\psi_{cc}=0.375}$ V, which corresponds to the inflection point of function ${I(\psi_{cc})}$ in the interval ${E_{1}(0)<q\psi_{cc}<E_{1}(0)+\hbar \omega_{op}}$. 
Fig.~\ref{fig:6} illustrates that the results obtained in the modified model A manifest better agreement with BTE calculations than those obtained in the modified model B.

To test validity of the modified  models A and B, we compare the simulation of transfer characteristics with experimental data obtained for CNTFETs with titanium (Ti) contacts and heavily doped CNT~\cite{Appenzeller2004}.
At the titanium-CNT contacts, SBs are formed. However,  CNT doping leads to thinning of the SBs, which improves the coupling between the electron reservoirs in source and drain~\cite{Appenzeller2004}. 
Hence, the experimental data should be similar to the I-V characteristics obtained in the modified model A developed for the ohmic metal-CNT contacts.

\begin{figure}[!t]
\centering
\subfloat{
\includegraphics{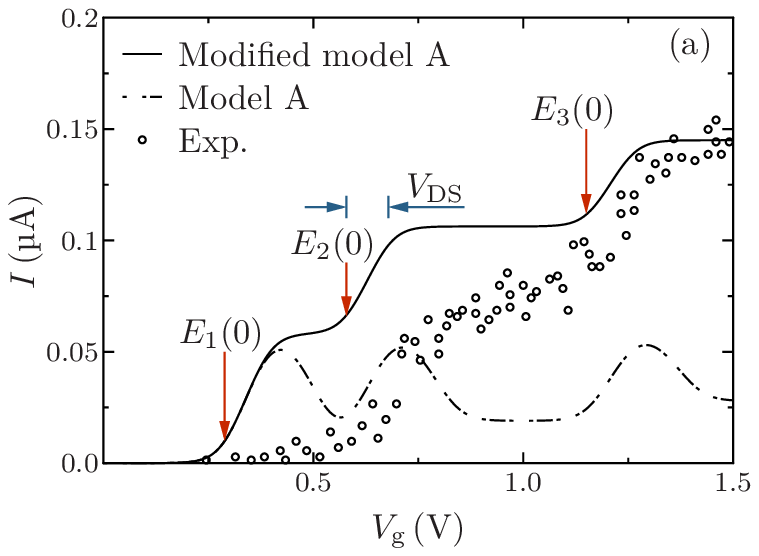}
\label{Exp_CM1}
}
\\
\subfloat{
\includegraphics{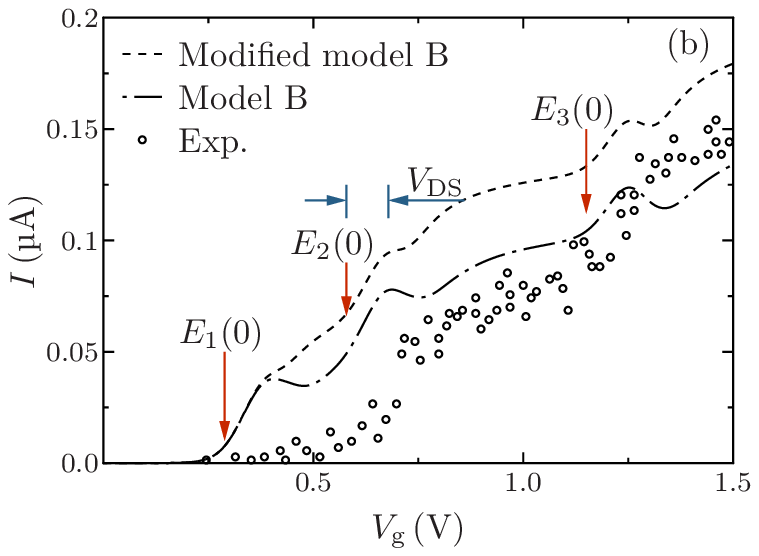}
\label{Exp_CM2}
}
\caption{(Color online) The source-drain current ${I}$ measured (data are from~\cite{Appenzeller2004}) and calculated (a) in the modified model A and (b) in the modified model B as a function of gate voltage at drain-source voltage ${V_{DS}=0.1}$ V. CNT chirality (19,0), gate length ${L_{g}=300}$ nm, temperature ${T=300}$ K, (a) ${\Delta=0.1V_{DS}}$, (b) ${\Delta=0.75k_{B}T/q}$. The electron subband edges are ${E_{1}(0)=0.289}$ eV, ${E_{2}(0)=0.579}$ eV, and ${E_{3}(0)=1.15}$ eV.}
\label{Exp_CM}
\end{figure}

Fig.~\ref{Exp_CM} shows the transfer characteristics obtained in the modified models A and B in QCL compared to the experimental data for CNTFET with a gate length of \SI{300}{\nano\meter}.
The simulated values of current ${I}$ are greater than the measured ones, because  the contacts resistance, electron scattering by impurities, surface polar phonons, and radial breathing mode phonons are not taken into account in our calculations.
For the benefit of comparison, the value of simulated current ${I}$  was scaled down by a factor of 0.008.
The drain-source voltage, ${V_{DS}=0.1}$ V, is less than the energy of optical phonons, ${\hbar \omega_{op} \approx \SI{0.16}{\eV}}$, therefore, the electrons are scattered only by acoustic phonons in this case.
The best agreement between the experimental data and simulation results is obtained in the framework of the modified model A (see Fig.~\ref{Exp_CM1}).
Both the measured and calculated I-V characteristics manifest a step-like dependence on the gate voltage.  
This is due to a saturation of the current corresponding to different electron subbands~ \cite{Appenzeller2004}.
Under the applied gate electric field, the current starts to sharply increase as one of the electron subband edges, $E_{m}(0)$, approaches the chemical potential of the source contact and electrons are injected into the CNT channel from the source. 
As soon as the electron subband edge approaches the chemical potential of the drain contact at ${q V_{g} \approx E_{m}(0)+qV_{DS}}$, electrons are injected into the channel from the drain and the net current saturates.
Fig.~\ref{Exp_CM1} indicates that the measured plateau widths are in agreement with the calculated values.
Although the results obtained in the modified model B agree with the experimental data better than those calculated by using the original model B, the small dip in the ${I(V_g)}$ dependence at high gate voltage indicates that the modified model B faces difficulties if many electron subbands are included (see Fig.~\ref{Exp_CM2}).    

\section{Conclusions}
\label{sec:level8}  

We have revised the model of electron back scattering  by emitting optical phonons in CNTFET.
We have shown, that the accumulation of the back scattered electrons at the bottom of the conduction subband affects the injection of electrons into the CNT channel from the drain contact and modifies the electron transmission probability.
Taking into account the new physical model, the limitations of models suggested in the literature have been overcome.
This allows to evaluate the compact models at high bias voltages needed especially for analog high-frequency circuit design.
The modified model A includes only analytical expressions and it is free of summation as needed for the evaluation of the modified model B.

\section*{Acknowledgment}

The authors would like to thank Prof. Schr\"oter for valuable discussions. Also, we wish to acknowledge Mr. Mothes for providing us with the BTE simulation tool.

\ifCLASSOPTIONcaptionsoff
  \newpage
\fi



\bibliographystyle{IEEEtran}
%
%
%

%

\begin{IEEEbiographynophoto}{Igor Bejenari}
 (PhD'2006) photograph and biography not available at the time of publication.
\end{IEEEbiographynophoto}

\begin{IEEEbiographynophoto}{Martin Claus}
 (Dr-Ing'2011) photograph and biography not available at the time of publication.
\end{IEEEbiographynophoto}





\end{document}